\documentclass[12pt]{article}
\usepackage[a4paper, left=2.5cm, right=2.5cm, top=2.5cm, bottom=2.5cm]{geometry} 
\usepackage{amsmath,amssymb,graphicx}

\begin{document}

\title{Metabolic scaling from Fibonacci dynamics} 
\author{
Dorilson Silva Cambui\\
Governo de Mato Grosso, Secretaria de Estado de Educação\\
Cuiabá, Mato Grosso, Brazil\\
\texttt{dorilson.cambui@edu.mt.gov.br}\\
\texttt{dcambui@fisica.ufmt.br}
}

\maketitle

\begin{abstract}
\noindent	
We propose a discrete model to determine the metabolic scaling exponent based on Fibonacci growth patterns and discrete biological development phases. In contrast to continuous fractal models such as the West-Brown-Enquist (WBE) theory, the present approach describes metabolic scaling as the cumulative result of successive discrete stages, each incrementally contributing to metabolic activity. The scaling exponent $b(n)$ emerges naturally from the logarithmic relationship between consecutive Fibonacci numbers, varying systematically with the organism's developmental stage. A refined logarithmic formulation significantly enhances quantitative agreement with empirical metabolic data across various mammalian species. This discrete framework effectively captures deviations from classical scaling laws, directly connecting recursive hierarchical structures with metabolic processes. Our model provides an alternative to traditional fractal transport approaches and can be naturally extended to hierarchical physical systems, opening new avenues to explore stage-dependent scaling phenomena in complex adaptive systems.
\end{abstract}

\section{Introduction}\label{sec1}

Metabolic rate represents the energy flux necessary to sustain physiological activity, quantified as the total energy expended per unit time by an organism under basal or active conditions. In biological scaling studies, metabolic rate $B$ is commonly described by a power-law function of body mass, expressed mathematically as $B = B_0 \cdot M^b$, where $b$ is the metabolic scaling exponent, $M$ denotes the organism's total mass, and $B_0$ is a normalization constant representing the metabolic rate of a reference organism with unit mass~\cite{west1997general}. Typically, this exponent is determined through linear regression analysis of metabolic rate data across species~\cite{kleiber1932body}, characterizing how energy requirements scale systematically with organism size.

The West-Brown-Enquist (WBE) model explains this phenomenon by considering fractal-like transport networks, predicting a nearly universal scaling exponent of $b \approx 3/4$ across diverse taxa~\cite{west1997general}. Although mathematically elegant and widely recognized, the WBE framework frequently fails to account for systematic deviations observed empirically, particularly during distinct developmental stages or under conditions of metabolic stress~\cite{savage2004effects}.

Recent studies highlight that biological transport networks are not purely continuous fractal structures but rather exhibit discrete, hierarchical, and self-similar organization optimized for resource distribution at multiple scales~\cite{banavar1999size}. Geometric patterns strongly correlated with the Fibonacci sequence and the golden ratio ($\phi = \frac{1+\sqrt{5}}{2}$) frequently emerge in biological systems, notably in phyllotaxis, vascular branching, and morphogenetic processes~\cite{Thompson,Prusinkiewicz}. These observations suggest underlying arithmetic constraints that guide developmental organization. Inspired by these recurring Fibonacci-based patterns, we propose a discrete model of metabolic scaling governed explicitly by Fibonacci recursion dynamics.

In our approach, an organism's total mass at developmental stage $n$, denoted as $M(n)$, is idealized as proportional to the Fibonacci term $F_n$, such that $M(n) \sim F_n$. Conversely, the metabolically active fraction corresponds specifically to the preceding term $F_{n-1}$. Consequently, the metabolic rate $B(n)$ is assumed to reflect metabolic contributions from all structures formed up to stage $n-1$, capturing the physiological reality that even seemingly inactive tissues (e.g., bones or lipid stores) play essential roles in basal metabolic functions~\cite{mccue2010starvation}. The relationship between the metabolically active fraction and the total mass at stage $n$ naturally defines a stage-dependent scaling exponent, $b(n)$, whose analytical form is derived explicitly later in this article. Our discrete formulation complements and extends the WBE model by capturing stage-specific metabolic scaling variations more accurately. Indeed, a refined logarithmic expression of $b(n)$ demonstrates enhanced quantitative agreement (up to 12\%) with empirical data across nine mammalian species (see Tables~\ref{tab1} and~\ref{tab2}), clarifying deviations from classical allometric predictions and quantifying the influence of birth mass $M_0$ on metabolic scaling.

Therefore, our discrete and recursive model emerges as a valuable alternative to continuous fractal approaches, providing new understandings of biological scaling. Its implications extend to the understanding of metabolic process and the development of optimized hierarchical systems. In the subsequent sections, we derive the explicit analytical form of $b(n)$, empirically validate the model, and discuss its broader conceptual and practical relevance.

\section{Model Formulation}

The formulation begins with the recursive identity of the Fibonacci sequence, $F_n = F_{n-1} + F_{n-2}$, which parallels a biologically motivated assumption for mass accumulation through discrete developmental stages, described mathematically as $M_n = M_{n-1} + M_{n-2}$. This recursion leads asymptotically to a geometric growth proportional to powers of the golden ratio $\phi$:
\begin{equation}\label{Mn}
	M(n) \sim M_0 \cdot \phi^n,
\end{equation}
where $M_0$ denotes birth mass, and $M(n)$ is the body mass at developmental stage $n$. The use of $\phi$ is justified by its recurrent emergence in efficient spatial arrangements, vascular branching architectures, and resource allocation patterns in diverse biological systems~\cite{jean1994phyllotaxis, meinhardt1982models}. Although actual biological growth varies due to environmental and physiological influences, Eq.~(\ref{Mn}) effectively represents a dominant scaling behavior inherent in hierarchical biological organization. Thus, it provides a useful abstraction rather than a literal description of organismal growth. Indeed, numerous biological phenomena empirically support such idealized Fibonacci-like patterns: phyllotaxis arrangements in plants exhibit spiral counts and divergence angles governed by Fibonacci numbers~\cite{douady1992}, and animal vascular or respiratory systems frequently show structural optimization consistent with $\phi$~\cite{katifori}.

Solving Eq.~(\ref{Mn}) explicitly for the growth stage yields
\begin{equation}\label{stage}
	n = \log_\phi\left(\frac{M}{M_0}\right),
\end{equation}
establishing a fundamental mapping between body mass and developmental stage. In mammals, these discrete stages can naturally correspond to infancy, juvenile, and adult maturity phases, each incrementally enhancing structural complexity and metabolic activity.

Unlike the classical power-law scaling relation $B = B_0 M^b$, in which $b$ is typically constant and empirically determined, our model introduces a discrete, stage-dependent exponent directly into the scaling relation:
\begin{equation}
	B(n) = B_0 M^{b(n)},
\end{equation}
thereby explicitly linking hierarchical structural growth to metabolic rate via Fibonacci recursion. By associating metabolically active mass to the Fibonacci term $F_{n-1}$ and total body mass to $F_n$, the scaling exponent becomes explicitly stage-dependent:
\begin{equation}\label{bn}
	b(n) = \frac{\log F_{n-1}}{\log F_n} \approx \frac{n-1}{n},
\end{equation}
valid under the asymptotic form $F_n \sim \frac{\phi^n}{\sqrt{5}}$. However, this approximation improves significantly for larger stages ($n \gg 1$), where the term $\log\sqrt{5}$ becomes negligible. Recognizing the limitations of this simplification, particularly for smaller values of $n$, we propose a refined logarithmic formulation:
\begin{equation}\label{bn2}
	b(n) = \frac{(n-1)\log \phi - \log \sqrt{5}}{n \log \phi - \log \sqrt{5}},
\end{equation}
which substantially enhances empirical alignment, especially at earlier growth stages.

Our discrete framework explicitly incorporates local quantized growth transitions, providing a biologically grounded explanation for empirical deviations from continuous fractal models such as the WBE theory. Particularly during early developmental phases, our refined expression accurately captures scaling behaviors observed empirically, displaying improvements over classical predictions up to about 12\%, and closely matching empirical values within the observed biological range ($0.686 \leq b \leq 0.870$)~\cite{white2005allometric}. 

\section{Results and Discussion}

To evaluate the robustness of our proposed scaling framework, we compared both simplified and refined versions of the model derived scaling exponent $b(n)$ with observed metabolic scaling across several mammalian species. Growth stages $n$ were computed from empirical birth mass ($M_0$) and adult mass ($M$) data, using the logarithmic relationship defined by Eq.~(\ref{stage}). Values for $M_0$ and $M$ were obtained from established biological databases and previous empirical studies~\cite{smith2003,gillooly2001,pantheria2009}, complemented by additional data sources~\cite{R1,R2,R22,R3,R4,R44,R5,R7,R9,R10,R11}, ensuring the representativeness and consistency of our dataset.

Our analysis reveals that the refined form of the scaling exponent (see Equation~\ref{bn2}),
consistently yields values within the empirically observed range $0.686 \leq b \leq 0.870$~\cite{white2005allometric}.
According to White et al.~\cite{white2005allometric}, this interval reflects physiological variability across basal and active metabolic states.
While the simplified approximation $b(n) = \frac{n-1}{n}$ broadly aligns with empirical observations, it tends to overestimate metabolic exponents, particularly at lower developmental stages ($n$ small). Inclusion of the logarithmic correction significantly improves quantitative agreement across all evaluated species and developmental stages. Table~\ref{tab1} presents calculated growth stages $n$, highlighting variability across species ranging from approximately 5.8 (mouse) to 9.2 (elephant). Table~\ref{tab2} further emphasizes that the refined formulation reduces deviations from empirical observations by an average of approximately 12\% relative to the WBE baseline ($b=0.75$).

\begin{table}[t]
	\centering
	\caption{{Calculation of parameter $n$.}
		Masses are given in kilograms and represent the lower and upper limits reported for each species at birth $M_0$ and adulthood $M$. The values of $n = \log_\phi(M / M_0)$ are calculated for these bounds, resulting in a corresponding range for each.}
	\label{tab1} 
	\begin{tabular}{lccc} 
		\\
		\hline
		\textbf{Species} &  {Range of mass $M_0$}  &  {Range of mass $M$} & $n = \log_\phi(M / M_0)$ \\
		\hline
		Rabbit & 0.03 - 0.08 & 3.5 - 5.5 & 7.852 - 9.890 \\
		Rat & 0.005 - 0.007  & 0.3 - 0.5 &7.809 - 8.508 \\
		Mouse & 0.001 - 0.0015  & 0.025 - 0.04 &5.846 - 6.689 \\
		Cat & 0.09 - 0.11   & 3.5 - 5 &7.190 - 7.607 \\
		Dog (medium)  & 0.3 - 0.5 & 25 - 35 &8.129 - 9.191\\
		Horse  & 45 - 60 & 450 - 600 &4.187 - 4.785\\
		Cow  & 30 - 50  & 600 - 800 &5.164 - 6.225  \\
		Elephant  & 95 - 120 & 4000 - 6300 &7.287 - 7.772\\
		Blue Whale    & 2000 - 3000 & 100000 - 150000 &7.287 - 8.130 \\
		\hline
	\end{tabular}
\end{table}

\begin{table}[t]
	\centering
	\caption{{The simplified and refined versions of the exponent $b(n)$ are calculated.}
		The refined Fibonacci-based exponent $b(n)$ is compared with the constant value $b = 0.75$ from the WBE model. The final column shows the relative deviation from the WBE prediction.}
	\label{tab2} 
	\begin{tabular}{lccc} 
		\\
		\hline
		\textbf{Species} & $b(n)$ simplified \hspace{0.1cm} \hspace{0.1cm} & \hspace{0.5cm} $b(n)$ refined \hspace{0.5cm} &  Rel. dev. (\%) \\
		\hline
		Rabbit & 0.873 - 0.899 & 0.838 - 0.878 & 11.758\\
		Rat & 0.872 - 0.882 & 0.837 - 0.854 & 11.607  \\
		Mouse & 0.829 - 0.851 & 0.760 - 0.801 & 1.391  \\
		Cat & 0.861 - 0.869 & 0.819 - 0.831 & 9.170 \\
		Dog (medium)  & 0.877 - 0.891 & 0.845 - 0.867 & 12.685  \\
		Horse  & 0.761 - 0.791 & 0.602 - 0.679 & 9.502 \\
		Cow  & 0.806 - 0.839 & 0.714 - 0.780 & 4.049  \\
		Elephant& 0.863 - 0.871& 0.822 - 0.836 & 9.586  \\
		Blue Whale    & 0.863 - 0.877 & 0.822 - 0.845  & 9.586 \\
		\hline
	\end{tabular}
\end{table}

Unlike the constant exponent predicted by the WBE model, our stage-dependent scaling model naturally evidences why early developmental stages exhibit distinctly lower exponents (e.g., $b(n)$\textit{refined} $ = 0.602 - 0.679$ for $n < 7$), while later stages ($n\gtrsim 7$) approach higher values, above $0.8$. The refined formulation yields an average exponent $\langle b(n)\rangle \approx 0.784$, compared to the simplified form $\langle b(n)\rangle \approx 0.845$, demonstrating not only analytical consistency but also clear biological realism.

\begin{figure}[h!]
	\centering
	\includegraphics[width=\linewidth, trim=10 5 8 5, clip]{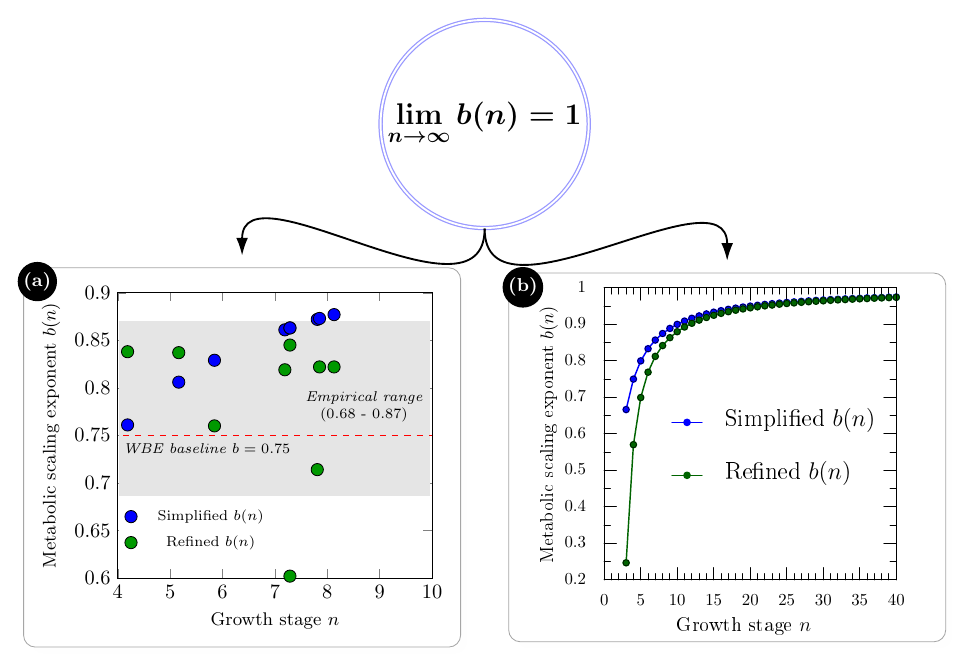}
	\caption{a) Model-derived $b(n)$ values with the empirical range and the WBE baseline ($b = 0.75$) highlighted, and b) the theoretical $b(n)$ curve showing its asymptotic approach to unity.}	
	\label{fig2}
\end{figure}

Both formulations asymptotically approach unity as the developmental stage $n$ increases, aligning with the biological intuition that, in mature organisms, further structural growth (such as new tissues or transport networks) adds progressively less to overall metabolic rate. This asymptotic behavior, evident in the theoretical curve shown in Figure~\ref{fig2}, reflects the model’s prediction that metabolic growth stabilizes in mature organisms, a feature not captured by fixed-exponent or continuous fractal models.

Figure~\ref{fig2} illustrates, on the left, the refined and simplified $b(n)$ values obtained from the model, based on the data from Tables~\ref{tab1} and~\ref{tab2}, considering only the lower bounds of $n$ and $b(n)$ for each species. Approximately 67\% of the simplified $b(n)$ values and about 89\% of the refined $b(n)$ values fall within the shaded gray area, which represents the empirically observed range for the metabolic exponent $b$~\cite{white2005allometric}. The red dashed line corresponds to the value $b = 0.75$ predicted by the WBE model. On the right, Figure~\ref{fig2} illustrates the theoretical $b(n)$ curve computed from Eqs.~\ref{bn} and~\ref{bn2}. The curve increases monotonically with $n$ and approaches unity asymptotically ($\displaystyle \lim\limits_{n\to\infty} b(n) = 1$). For very early stages with $n \leq 4$, the refined expression fails by producing values that are too low for the metabolic exponent. Concretely, $b(2) \approx -2.05$, $b(3) \approx 0.25$ and $b(4) \approx 0.57$, all lying at or below the empirical lower bound. This behavior indicates a limitation of the refined formulation at the onset of the recursive architecture and suggests that its reliable domain begins only after the earliest developmental phases, consistent with $n > 4$ in our dataset.

Finally, recognizing biological variability and environmental factors not explicitly incorporated into our model, future validations should consider broader taxonomic groups and distinct ecological contexts. Nevertheless, our discrete, Fibonacci-based approach provides an effective analytical tool for understanding stage-dependent metabolic scaling, opening pathways for further exploration in both biological and engineered hierarchical systems.

\section{Conclusion}

We have introduced a novel discrete, recursive framework for metabolic scaling theory based explicitly on Fibonacci growth dynamics, offering a understandings shift from traditional continuous fractal assumptions. This approach addresses discrepancies between classical scaling predictions and empirical biological data by defining the metabolic scaling exponent $b(n)$ explicitly as a function of developmental stage $n$, rather than assuming it as a constant.

By linking mass accumulation directly to discrete, recursively structured growth layers, our model predicts systematically varying scaling exponents that asymptotically approach unity as organisms mature. While the simplified expression $b(n) = \frac{n-1}{n}$ provides analytical simplicity, the refined logarithmic formulation significantly improves agreement with empirical data, consistently achieving an average relative deviation of 8.82\% (minimum 1.39\%, maximum 12.69\%) across the mammalian species analyzed. Particularly noteworthy is the model’s ability to accommodate both large deviations from the classical WBE prediction, such as in the dog (12.69\%) and horse (9.50\%), and very small ones, as in the mouse (1.39\%).

Although Fibonacci patterns have been traditionally explored in spatial morphogenesis and phyllotaxis~\cite{douady1992}, our findings highlight their functional role in metabolic scaling through recursive architecture. The model is especially relevant for multicellular organisms characterized by hierarchical developmental progression, though it may have limitations when applied to systems lacking clearly defined structural growth phases.

By establishing a clear mechanistic connection between developmental progression and metabolic scaling, our approach integrates Fibonacci recursion into metabolic theory, providing new conceptual perspectives and opportunities for exploring hierarchical optimization strategies in biological, physical, and engineered systems.

Although the assumption of idealized growth represented by $M(n) \sim M_0 \cdot \phi^n$  might initially appear speculative, its demonstrated predictive accuracy across empirical datasets underscores its utility and robustness. Far from a mere mathematical curiosity, this discrete recursive framework emerges as a viable and effective alternative to conventional scaling models, offering practical tools to infer metabolic rates directly from measurable developmental parameters.

\end{document}